# Managing Changes in Citizen-Centric Healthcare Service Platform using High Level Petri Net

Sabri MTIBAA
LI3 Laboratory / University of Manouba
National School of Computer Sciences
2010 Manouba, Tunisia
Sabri.Mtibaa@gmail.com

Moncef TAGINA
LI3 Laboratory / University of Manouba
National School of Computer Sciences
2010 Manouba, Tunisia
Moncef.Tagina@ensi.rnu.tn

*Abstract*— The healthcare organizations are facing a number of daunting challenges pushing systems to deal with requirements changes and benefit from modern technologies and telecom capabilities. Systems evolution through extension of the existing information technology infrastructure becomes one of the most challenging aspects of healthcare and the adaptation to changes is a must. The paper presents a change management framework for a citizen-centric healthcare service platform. A combination between Petri nets model to handle changes and reconfigurable Petri nets model to react to these changes are introduced to fulfill healthcare goals. Thanks to this management framework model, consistency and correctness of a healthcare processes in the presence of frequent changes can be checked and guaranteed.

I. INTRODUCTION

The traditional method of receiving healthcare required a patient to visit their doctor's office, or a hospital; now doctors, hospitals, and healthcare ecosystems are increasingly brought to the patient. The baby boomer generation is entering a new stage of their life; they are adopting and demanding better delivery models for quality and access to care [1]. This has caused a paradigm shift in the healthcare system.

The new culture is patient centered utilizing care coordination that is focused on successful outcomes that depend on new innovations, technology and meaningful data, including collection, delivery, ease of use, intelligence, and reporting [2]. As human life expectancy continues to increase and aging populations lead to higher health treatment costs, telehealth is on the top of the regulators agenda.

The focus on the climbing cost to deliver and maintain quality healthcare is no longer in the peripheral view, but a clear line of sight for the patients, healthcare providers, regulators, and payers. This brings new requirements for healthcare professionals to share information, communicate and collaborate in real time from multiple locations, because medicine is a collaborative science.

Communications become a strategic asset for a strongly needed healthcare transformation technology. It must be deployed to this field in order to ensure better context for medical decisions, reduce administrative costs and improve patient safety by reducing errors. The healthcare community has recognized the need to transform from the current hospital centralized treatment-based mode to prevention-oriented comprehensive healthcare mode in which hospitals, communities, families and individuals are closely involved. The new mode needs to provide individuals with intelligent health information management and healthcare services. It allows them to enjoy medical prevention and healthcare services in their daily life.

The advancement of information technology (IT) brings more opportunities for innovations in the healthcare area. The use of service oriented technologies such as SOA, Web Services (WS) allows service providers to reduce and simplify integration process, to abstract network capabilities (e.g., call control, presence, location, etc.), and create personalized and blended services (both internally and with 3rd party partners) [3]. These technologies facilitate the construction of service systems with higher reusability, flexibility, extensibility, and robustness.

Cloud computing is evolving as an important IT service platform with its benefits of cost effectiveness and global access. Built upon Enterprise Service Bus (ESB) as an integration backbone [4], this paper presents a novel citizen-centric healthcare service platform. One of the much-touted potentials of this platform is the ability to construct healthcare composite Web services on demand, relieving telecom operators from the intricate details of how technologies work so they can focus on the business aspects.

As healthcare services aggregated in the proposed healthcare service platform increases, the complexity of managing changes will grow and the manual management of changes becomes not practical [5]. One of the greatest promises of the healthcare platform is ability to self-adapting to guarantee goals achieving. Abstracting changes in business relationships claim a framework to manage changes without any impact. For instance, the proposed healthcare platform needs to achieve the plug-in/plug-out Web services with little overhead while guaranteeing properties. These properties can be classified either *functional* or *non-functional*. Functional properties address the functionalities that healthcare platform have to fulfill. The non-functional properties refer to events surrounding the functional properties.

Change management is a critical component in the deployment of healthcare platform. We identify two main approaches dealing with changes: top-down and button-up [6]. A top-down approach focuses on changes that are usually business mandated. These changes are motivated by the business goal, and do not consider the uncertainty of the underlying member services. The second type of changes is referred to as bottom-up changes because Web service providers are the initiators of changes. Bottom-up changes are initiated by the member services. These changes are initiated in the Web service environment, and eventually translate into top-down changes. A service operation may become unavailable and trigger dependencies services and users in order to replace this service. In this paper, we will concentrate on this aspect.

In this paper, we present a conceptual module for management of bottom-up changes using Petri nets. In our work, we use Petri nets to model handling and adaptive changes in healthcare platform. We model changes using Petri nets because of their applicability to a Web services composition modelling.

The behaviour of a composite Web services is described by the evolution of its Petri net model [7]. As the Petri net evolves, the system goes through different safe and unsafe states that can be completely defined by the marking of a Petri net model. Furthermore, Petri nets map directly to our change specification. They also preserve all the details of our change specification while modelling the changes accurately. For instance, Petri nets can easily represent the safe an unsafe states of web services composition. They represent changes between these states as transitions. In addition, the use of reconfigurable Petri nets allows us to incorporate our mapping rules into the Petri net model. This allows us to completely model our change specification.

The remainder of this paper is organized as follows. In Section 2, we use the healthcare platform architecture of and a scenario from this domain to motivate our work. It will also be used as a running example. Section 3 presents a bottom-up specification of changes. In Section 4, we describe our change management model which is based on Petri nets. Finally, we conclude in section 5.

## II. HEATHCARE SERVICE PLATFORM ARCHITECTURE

In this section, we present the global context of our work and an overview about service oriented architecture, cloud computing and enterprise service bus. Then, the healthcare services platform architecture is exposed and some basic concepts and definitions are explained.

### A. Context and Background

Below we have summarized a few key notions and technologies that should be of significant value to the design of healthcare architecture.

*1) Service Oriented Architecture (SOA):* In this IT architecture, applications and more discrete software functions are network-based, loosely coupled and available on demand to authorized users or to other applications or services. Although SOA is not a new concept, the emergence of Web services as a standard way to expose, describe, access and combine services has given new life to this approach to computing.

The key idea of SOA is the following: a service provider publishes services in a service registry [7]. The service requester searches for a service in the registry. He finds one or more by browsing or querying the registry. The service requester uses the service description to bind service. These ideas are shown in Fig. 1.

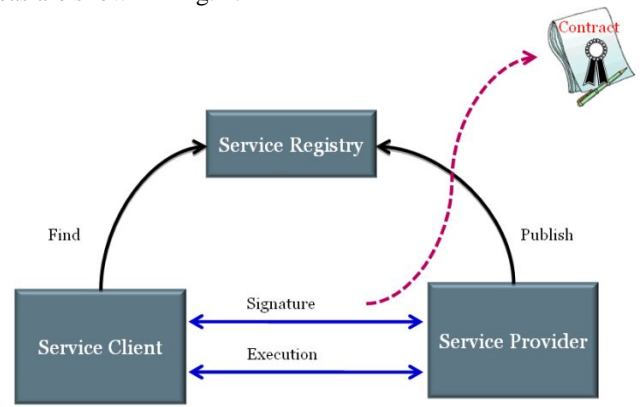

Fig. 1. Reference architecture of web services- SOA

*2) Cloud Computing:* Cloud computing called also *utility computing* refers to an IT service model and platform that provides on-demand based IT services over the internet (see Fig. 2). The five essential characteristics are: on-demand self-service, broad network access, resource pooling, rapid elasticity, and measured service [8]. The three services models include:

- SaaS (Software as a Service) which delivers software service on demand, such as, salesforce.com – Customer Relationship Management (CRM) service and Google Gmail;

- PaaS (Platform as a Service) which provides the computing platform for companies to deploy and customize business applications on demand, such as, Google App Engine and Microsoft's Azure;

- IaaS (Infrastructure as a Service) which offers data center, infrastructure hardware and software resources on demand, such as, Amazon Elastic Compute Cloud (EC2) and VMware vCloud Datacenter. Both of these resources provide virtual computers for renters to run their business applications.

The four major deployment models include: private cloud, public cloud, community cloud, and hybrid cloud. Companies normally adopt different service models and deployment models depending on their unique business processes and demands on IT services.

Cloud computing today is an evolution and application of modern ICT including server virtualization, autonomic computing, grid computing, server farm, network storage, and web service.

*2) Enterprise Service Bus:*

ESB is one piece of an infrastructure that might help facilitating the implementation of a SOA, but it is not a perquisite. There are many aspects of an ESB that fit well with the SOA model, and denying its possible usefulness would be counterproductive, but the two are not completely inter-dependent [4].

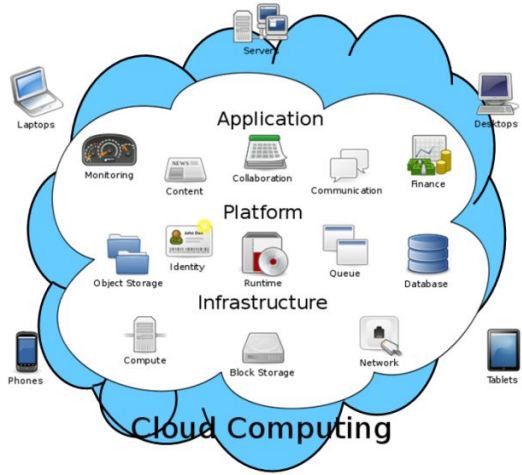

Fig. 2. Cloud computing architecture

Fig. 3 depicts the base functional elements within an ESB. It includes:

- Data transformation.
- Application adapters.
- Automation of processes.
- Transformation.
- Routing.
- Messaging.
- Event triggering.

If we consider some of these functional elements it can be seen that items such as application adapters fall neatly into the product category, while routing and messaging are more of an architectural consideration.

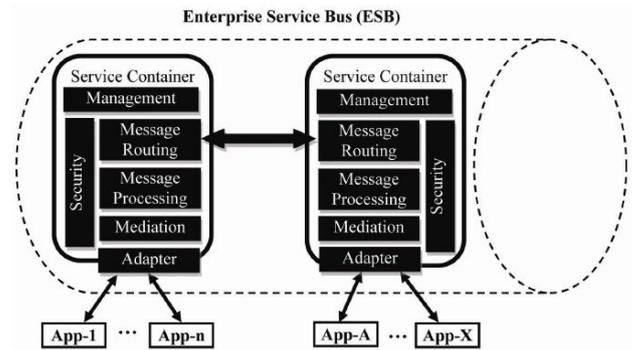

Fig. 3. ESB architecture

*B. System Architecture*

First, we present our Healthcare Service Platform (HSP). It intends to provide personalized healthcare services for the public. The healthcare value chain is complex. It consists not only of healthcare providers, but also of payers (government, employers and patients), fiscal intermediaries, distributors and producers of pharmaceuticals and devices [9].

The HSP does not attempt to address this complete value chain. It focuses on the delivery of healthcare services. It is an end-to-end reference architecture that focuses on meeting the needs of citizens, patients and professionals. Its architectural diagram is given in Fig. 4.

We distinguish three main components, i.e. body sensor networks (BSN), IaaS cloud, healthcare delivery environment.

- BSN: according to circumstances and personalized needs, appropriate health information collection terminals (i.e. sensors) are configured for different individuals. BSN is used to provide long term and continuous monitoring of patients under their natural physiological states. It performs the multi-mode acquisition, integration and real-time transmission of personal heath information anywhere [10].

- IaaS cloud: modern healthcare is information driven. Healthcare providers are making progress in building an integrated profile of patients. This data sits in systems throughout the enterprise including the HER and many other electronic systems throughout the enterprise and community [11]. This component achieves the rapid storage, management, retrieval, and analysis of massive heath data. It mainly includes *Electronic Medical Record* (EMR) repository. It considers also personal health data acquired from BSN.

- Healthcare delivery environment: it includes a personal health information management system. It replaces expensive in-patient acute care with preventative, chronic care, offers disease management and remote patient monitoring and ensures health education/wellness programs.

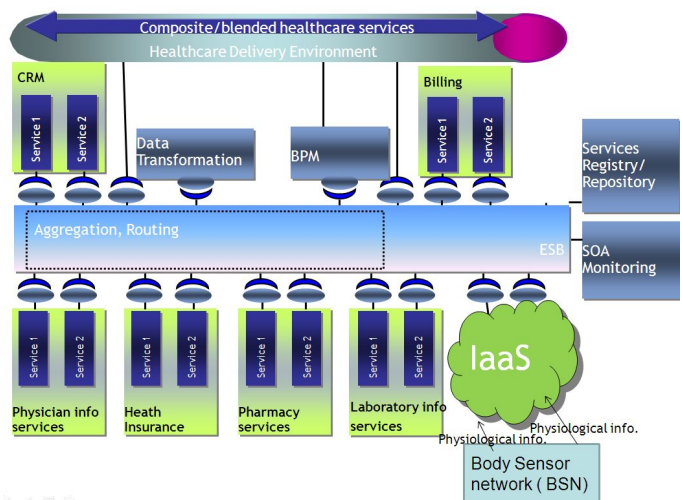

Fig. 4. HSP architecture

### C. Healthcare Web Services Provided by HSP

In PHISP, we adopt the design idea of SOA and Web service technology for its design and implementation. The majority of its functional modules are developed and packaged in the form of services [8]. Here, we overview some of them as follows.

- *PhysInfoWS*: this service can acquire some general physiological signals such as body temperature, blood pressure, and saturation of blood oxygen, electrocardiogram, and some special physiological signals according to different sensor deployment for different users. User's ID number is required.
- *EnvInfoWS*: for a unique ID number, this service can acquire temperature, humidity, air pressure and other environmental information for this user.
- *SubjFeelWS*: it can acquire the user subjective feelings, food intake, etc., and the information is often provided by the user from the terminal.
- *CoronaryDiagWS*: it can analyze the information according to a series of analysis models, which are built for coronary heart disease, and then produce preliminary diagnostic results.
- *AssessmentWS*: this service can assess the status of the patient's health risk based on the diagnostic results and the EMR information of the patient.
- *EmrWS*: this service can output the user's medical history information.
- *GeoWS*: it can return the user's location.
- *EmerWS*: it can raise an alarm to the user in case of illness.
- *GuideWS*: it can provide the patient with preventive measures especially items that need attention.

### D. Healthcare Service Scenario

A way to motivate and illustrate this work, we presents an example of healthcare service scenario. We distinguish three main layers: service, business and HSP. The service layer consists of available web services, and the business layer represents the Web service like operations typically ordered in a particular application domain. We refer to the selected services as member services (see Fig. 5).

Key healthcare environment objectives include:

- **Allowing people to stay in their homes to an older age.** By doing this, we can reduce the economic burden of dedicated care facilities and improve quality of life for a substantial proportion of the aging population.
- **Using televisions to keep in touch.** Another use of camera technology is in conjunction with an IPTV set-top box and connection back to a Contact Center.
- **Using wireless toys for always-on monitoring and communications**. The wireless home network itself enables a new class of device that has significant healthcare implications.

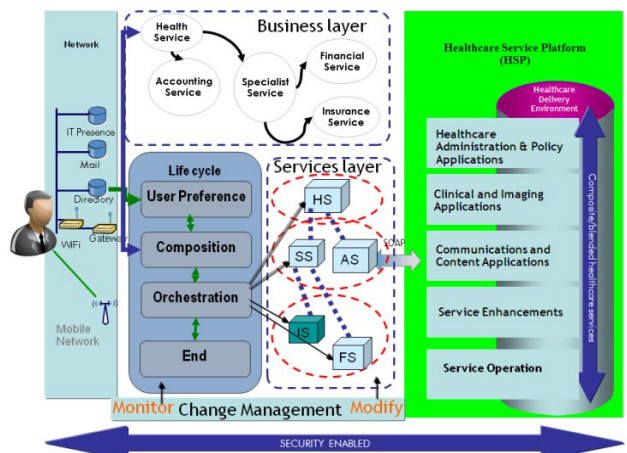

Fig. 5. Healthcare Service Scenario based on HSP

Let us assume that a *citizen* establishes a need for a business objective (healthcare service). Typically, he starts with formulating the business strategy (or goal). During the planning, some services can be identified: *HealthService*, *AccountingService*, *SpecialistService, FinancialServcie* and *InsuranceService*. Second, the *senior citizen* develops a specification listing the services to be composed through a graphical interface. We assume that *HS*, *AS*, *SS*, *FS* and *IS* are selected and orchestrated. The third step is the orchestration where member services that match the specified high level configuration are selected and invoked.

We describe here the ideal scenario: the *senior citizen* subscribe to *HealthService*. Then all information regarding who contacts it and when are forwarded to *AccountingService*. *HealthService* forwards also the received data to *SpecialistService* in charge of checking the received values. After analyzing the received values, the team sends a confirmation or an adjustment of the medication doses. The *FinancialService* and *InsuranceService* are executed to finalize the process.

*E. Modeling Healthcare process using High Level Petri Nets*

The modeling of healthcare process is as crucial as the implement of healthcare service platform. The formal representation of real healthcare process with Hierarchical Petri-Nets is easy to understand. Fig. 6 shows a Hierarchical Petri Net that describes our healthcare service scenario offered by HSP.

From the above, the first part of hierarchical healthcare process net N with refinable transition named 'Health Service is shown below. It is refined with the attachment of web services net N'. The planned web service is the assembly of the set of web services presented previously (*PhysInfoWS, EnvInfoWS, SubjFeelWS, CoronaryDiagWS, AssessmentWS, EmrWS, GeoWS, EmerWS, GuideWS*).

However some changes to member services may handle some inconsistency in the composition and orchestration. Each service layer change presents a functional and non-functional change that may happen in a member service.

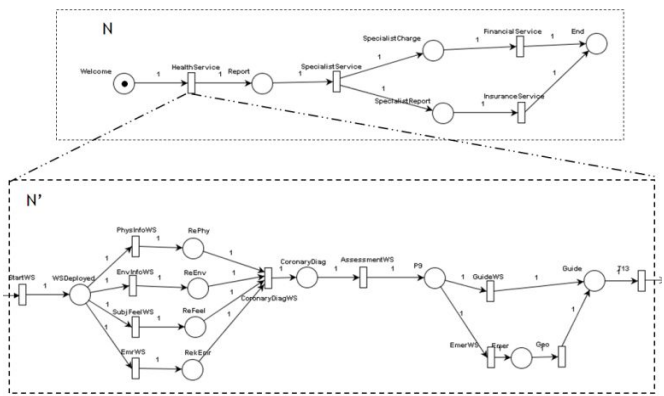

Fig. 6. Healthcare process modeled by hierarchical Petri nets

For example, in case of non availability of a *SpecialistService*, a change management is required to ensure that the healthcare system is remaining profitable.

### III.  CHANGE SPECIFICATION

Managing bottom-up changes is highly dependent on the services that compose the system. Therefore, it is quite important to define the changes that may occur to web services and then map them to into the system level.

In this section, we present bottom-up changes. We define the set of handling changes. Handling change ($\theta$) is defined for changes that occurred at the service level (for example Web service availability) while adaptive changes ($\Omega$) are related to changes at the business level (for instance the selection of alternative service).

*A. Changes overview*

For each change, a transition will be associated between two states: precondition and postcondition. In our scenario presented, a precondition for SS unavailability is that it was available and the postcondition is that it has switched to unavailable state.

Handling changes will be modeled using Petri nets. Our classification of triggering changes is based on the traditional approaches from the fields of software engineering and workflow systems [7]. A handling change is initiated at the service level such as the operations, the availability, etc. Therefore, we can distinguish several handling changes based on Web service properties.

The Web service properties can be sorted into two categories: functional and non-functional. Fig. 7 shows the handling changes: functional and non-functional.

- *Non-functional changes:* we assume that the non-functional parameters represent the dependability and response aspects associated with a member service. Service dependability can be set to two possible values (i.e., available or unavailable). Alternatively, service cost values may take more than two possible values.

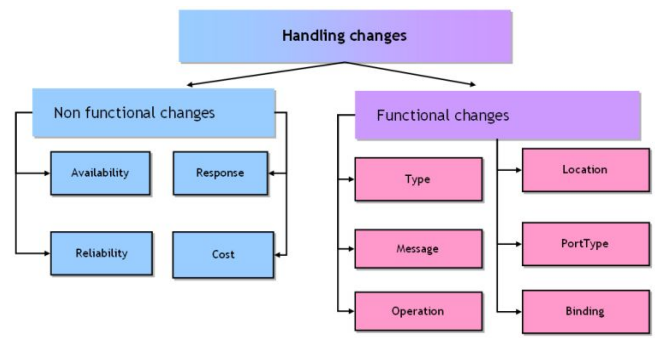

Fig. 7. Handling changes categories

- *Functional changes*: this category of changes is related to a service WSDL description [12]. We consider functional changes as a combined execution of elementary operations: *remove* and then *add*. We distinguish two different functional changes: structural and behavioral (see Fig. 6). Structural changes are related to the operational aspects of a Web service. For example, a structural change in a healthcare service can be a consequence of change in the operations offered to a citizen. Functional changes to a member Web service occur when its WSDL description is modified.

- *Adaptive changes:* Adaptive changes may occur at the composition and orchestration levels. Fig. 8 shows the adaptive changes considered in our model. In our scenario, when the healthcare system is interrupted by a change in *SS*, it reacts to the change after suspending execution. This may be accomplished by raising a fault, compensating for the change at the composition layer, and calling of an alternate service.

For example, if *SS* becomes unavailable, business layer will be search for equivalent service to continue execution to ensure that there is no high level impact on user demands.

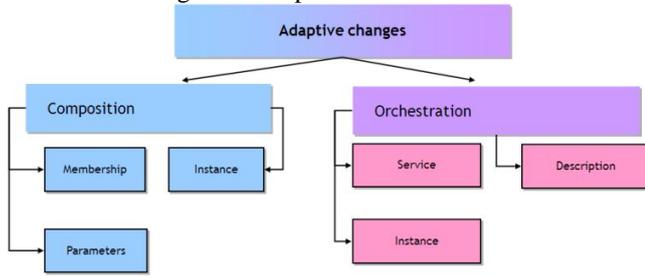

Fig. 8. Handling changes categories

Now, we will discuss the impact of $\theta$ changes to the healthcare business layer. A mapping details how change instances in one layer correspond to changes in another layer. These mapping must remain consistent in the presence of frequent changes. Handling changes have a reactive impact on the business layer. For instance, a $\theta$ change in availability maps to $\Omega$ change of change instance.

## I. CHANGE MODEL

In this section, we introduce a change model to accurately identify eventual types of changes in a composite Web services.

### A. Handling Changes Model using Petri Nets

Petri nets or PN are a well-founded process modeling techniques that have formal semantics. They have been used to model and analyze several types of processes including protocols, and business processes.

Visual representations provide a high-level, yet precise language, which allows reasoning about concepts at their natural level of abstraction. Services are basically a partially ordered set of changes. Therefore, it is a natural choice to map it into a Petri net. Moreover, the semantics delivered by Petri nets can be used to model the standard behavior of composite Web services described by BPEL [7].

We formalize the change model for triggering changes by introducing Petri-Net-Handle (PNH) which is defined as follows.
The algebraic structure of PNH = ($P$, $T$, $F$, $P_0$, $P_n$) if the following conditions hold:
- $F \subseteq (P \times T) \cup (T \times P)$
- $P \cap T = \emptyset$
- $P \cup T \neq \emptyset$
- $P_i \in P$
- $P_0 \in P$

where:
- $P$ is a finite set of places representing the states of Web service.
- $T$ is a finite set of transitions representing changes to Web service.
- $F$ is called the web services action flow.
- $P_0$ is the input place, or starting state of the Web service
- $P_n$ is the output place, or the ending state of the Web service

Fig. 9 represents the model of non-functional changes to Web services. It is composed from five places and four transitions. *PS* is the initial place of $PNH_N$. It represents the initial state of the Web Service. *PS* consists of four tokens, each representing one of the four non-functional changes. The token corresponding to change is fired each to represent dynamic evolution. If more than one change occurs, the corresponding token for each change type is fired.

For instance, if a member services (i.e Web service) becomes unavailable, the transition will be enabled and the corresponding token will be fired.

Table I. gives summary about non-functional changes.

TABLE I. NON FUNCTIONAL CHANGES

| Change | $\theta$ | Pre | Post |
|---|---|---|---|
| alterAvailibility | $\theta_A$ | PSA | PS'A |
| alterReliability | $\theta_R$ | PSR | PS'R |
| alterCost | $\theta_C$ | PSC | PS'C |
| alterResponsivenss | $\theta_{Re}$ | PSRe | PS'Re |

The subnet representing dependability changes in $PNH_d = (P_d, T_d, F_d, P_{0d}, P_{nd})$, where $P_d = \{PS, PS'Re, PS'A\}$ and $T_d = \{TRe, TA\}$. The place *PS* is corresponding to the state of available and reliable service. *PS'A* represents a service that becomes unavailable.

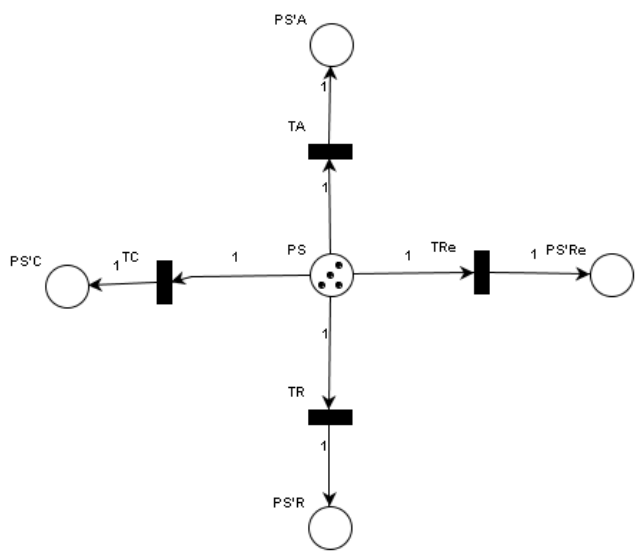

Fig. 9. Handling non-functional changes categories

When a service becomes unavailable, the token (representing the availability property) is moved from *PS* to *PS'A*. A similar behavior is observed when the service becomes unreliable. *TA* represents alterAvailibility, and *TRe* represents alterReliability.
The same logical structure can be applied for functional changes.

### B. Modeling Adaptive Changes with Reconfigurable Petri Nets

We have surveyed extensions of Petri nets for modeling reactive changes. Reconfigurable Petri-nets provide formalism for modeling these changes. It is a class of high level Petri nets.

They support internal and incremental description of changes over a uniform description. Reconfigurable Petri nets are an extension of Petri nets with local structural modifying rules performing the replacement of one of its subnets by other subnets [13]. The tokens in a deleted place are transferred to the created one.

We formalize the change model for adaptive changes by introducing PNAC (PNAC) which is defined as follows. The algebraic structure of PNAC= (P, T, F, R, I) where:

- $P = \{P_1 \ldots, P_n\}$ is a non empty and finite set of places
- $T = \{T_1, \ldots, T_n\}$ is a non empty and finite set of transitions disjoint from $P (P \bigcap T = \Phi)$
- $F: (P \times T) \bigcup (T \times P) \rightarrow$ IN is a weighted flow relation. A rewriting rule is a map $r : P_1 \rightarrow P_2$ whose domain and co domain are disjoint subsets of places $P, P_1 \subseteq P, P_1 \bigcap P = \Phi$
- $R = \{r_1, \ldots, r_n\}$ is a finite set of structure modifying structure rules.

- *I* represent the initial state: the first configuration of composition in business layer. The domain of *I* is $HCE_0$.

Table I. gives summary about adaptive changes.

TABLE II. ADAPTIVE CHANGES

| Change | $\Omega$ | *Pre* | *Post* |
|---|---|---|---|
| alterState | $\Omega_{ST}$ | VEST | VE'ST |
| alterServiceInstance | $\Omega_S$ | VES | VE'S |
| alterCost | $\Omega_C$ | VEC | VE'C |

We consider the scenario containing five places corresponding to adaptive changes:
- $HCE_S$ is the set of places $\{HCE_0, HCE_1, HCE_2, HCE_3, HCE_4\}$ where S represents alterState.
- $HCE_v$ is the set of places $\{HCE_5, HCE_6, HCE_7, HCE_8, HCE_9\}$ where V denotes alterServiceInstance.
- $HCE_w$ is the set of places $\{HCE_{10}, HCE_{11}, HCE_{12}, HCE_{13}, HCE_{14}\}$ where W denotes alterOrder.

Fig. 10 shows a PNAC representing initial statechange in service orchestration. The adaptive changes using Reconfigurable Petri net representing modification on service state (see Fig. 11), removal of service (see Fig. 12), and addition of service (see Fig. 13) are presented.

### C. Change Management Framework

We use our Petri net change specification as the basis for handling changes in our healthcare environment. The framework of change management is divided into two modules: detection and reaction.

After the change specification is defined, we begin the management. Detecting the respective changes is the first step of change management.

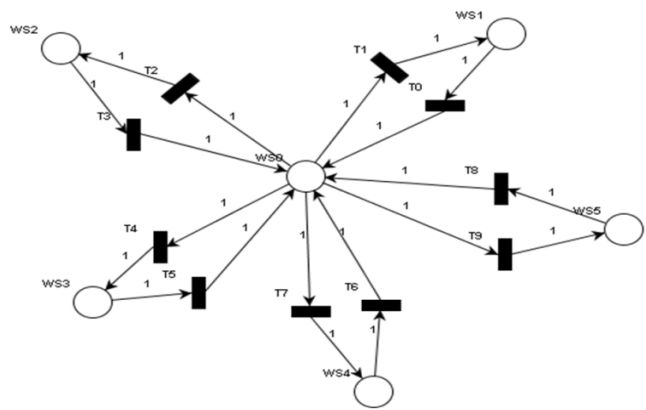

Fig. 10. Reconfigurable Petri nets for Reactive changes- initial state

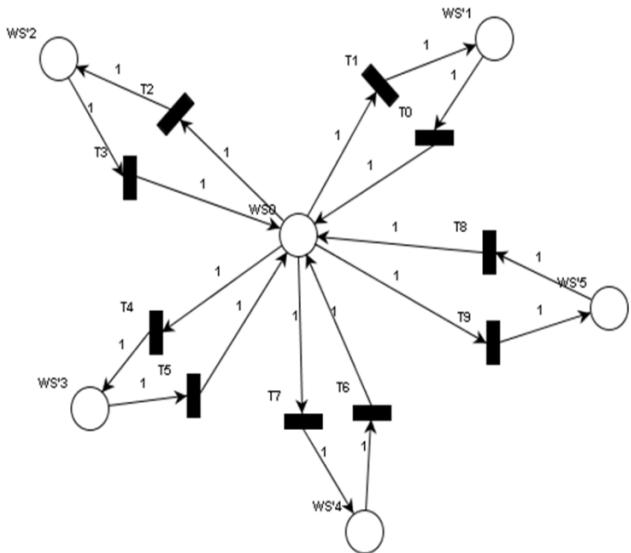

Fig. 11. Reconfigurable Petri nets for Reactive changes- after change

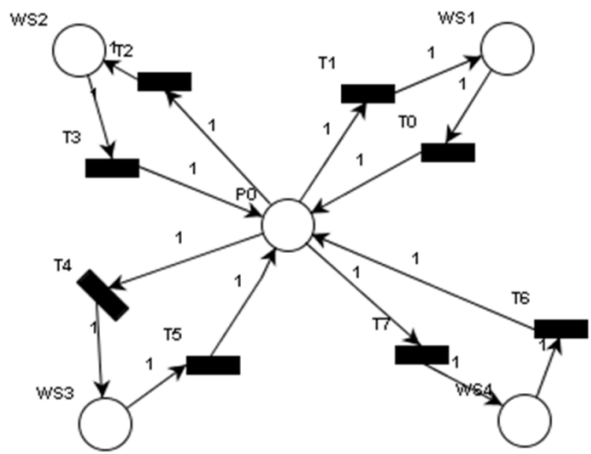

Fig. 12. Reconfigurable Petri nets for Reactive changes- after removal of service

All changes identified in the handling changes models are subject to detection. Detection involves an agent that monitors the Web service. Each change type has an associated set of rules for detection. For example, a *SpecialistService* may change the input parameters (i.e required information provided by patient), when this change occurs, the healthcare system must detect this change using some predefined detection rules.

Each detected change must be forwarded to monitoring service; and then the composition strategy must be updated. The notification and polling mechanism are mainly the techniques to awareness that a change has occurred. These techniques require that a monitoring service periodically send "Refresh" and "Alive" messages to detect unavailable services and also renew membership.

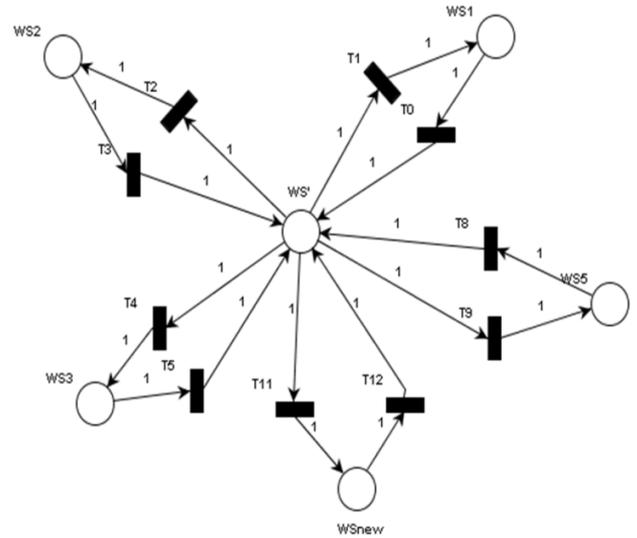

Fig. 13. Reconfigurable Petri nets for Reactive changes- after addition of new service

The changes are detected at the service layer and represented as an incidence matrix. Some rules are identified for detecting functional and non functional changes.

We define a rule for mapping the change into the defined Petri-net: first of all, the current service state is corresponded to a set of precondition places in the triggering Petri net and the updated service state as the set of postcondition places in the triggering Petri net. Then, a comparison between the values precondition and postcondition places of the Petri net is done. Depending on result retuned, a token is placed in the respective precondition place. This token will enable the change transition only if a difference is found.

Let us consider the example where the service WS service change availability (due to maintenance reasons) and the other attributes remain constant. In this case, we map the change into the non-functional Petri net. The service agent responsible of monitoring of WS will generate the incidence matrix corresponding to the Petri net model. This agent is in interaction with healthcare service platform to detect effective changes in the execution environment and map it into a Petri net model defined in this section ( for example, unavailability of a WS). The Centralized agent is the module that reacts to these changes and purpose a reconfiguration in the execution environment to guarantee consistency and correctness of the healthcare processes. Fig. 14 shows the different modules designed for management framework to detect and react to different changes in healthcare services.

Based on the information sent by service agent we define how to execute adaptive change. After receiving the matrix indicating the change that occurred, the handling change is mapped to the appropriate reactive change. We can list some considered reaction techniques in our system:

- In case of add, the newly service member will be considered. It can be taken into account in load balancing context or as back-up alternative.

- In case of unavailability of a service, if it is critical then the orchestration will be paused. Since a heartbeat is activated to check the status of the service; the orchestration will wait the service availability otherwise orchestration will exit (depending on configurable heartbeat number).

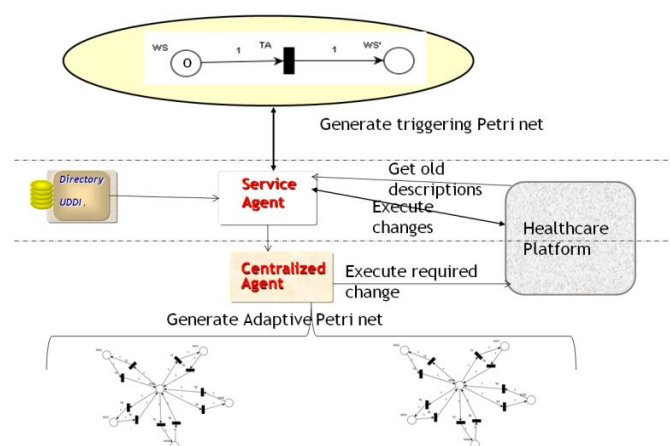

Fig. 14. Adaptive Petri net generator module in change management framework

## II. CONCLUSION AND FUTURE WORK

In this paper, we first presented a novel architecture of healthcare services platform. Second, we exposed the bottom-up approach focusing on handling changes that may occur in this system and then mapped to adaptive changes. We used a formal change model based on High Level Petri Nets to accurately represent these changes.

Future work includes an extension of change management framework. We plan to include a top-down approach to specifying changes. A full simulation prototype taking into account priority in changes and the estimation of their frequency based on measures represent actions planned to enhance actual healthcare platform.

AUTHORS PROFILE

**Sabri Mtibaa**

He is currently a Ph.D. student in the National School for Computer Sciences of Tunis, Tunisia (ENSI). He received the master degree from High School of Communication of Tunis, University of Carthage, Tunisia (Sup'Com) in 2008. His current research interest includes web service composition using Petri nets as well as system verification and QoS aware.

**Moncef Tagina**

He is a professor of Computer Science at the National School for Computer Sciences of Tunis, Tunisia (ENSI). He received the Ph.D. in Industrial Computer Science from Central School of Lille, France, in 1995. He heads research activities at LI3 Laboratory in Tunisia (Laboratoire d'Ingénierie Informatique Intelligente) on Metaheuristics, Diagnostic, Production, Scheduling and Robotics.